\documentclass[prd,floats,aps,showpacs]{revtex4}

\usepackage{dcolumn,epsfig}

\begin{document}

\title{Perturbative evolution of particle orbits  around Kerr
black holes: time domain calculation}

\author{Ram\'{o}n L\'{o}pez-Alem\'{a}n$^{1}$}

\author{Gaurav Khanna$^{2}$}

\author{Jorge Pullin$^{3}$}

\affiliation{$^{1}$Physical Sciences Department, University of
Puerto Rico - R\'{\i}o Piedras, San Juan, PR, 00931}

\affiliation{$^{2}$Natural Science Division,
Long Island University, Southampton, NY 11968 }

\affiliation{$^{3}$Department of Physics and Astronomy, Louisiana
State University, 202 Nicholson Hall, Baton Rouge, LA 70803-4001}

\begin{abstract}
We consider the problem of the gravitational waves produced by a
particle of negligible mass orbiting a Kerr black hole. We treat the
Teukolsky perturbation equation in the time domain numerically as a
2+1 partial differential equation.  We model the particle by smearing
the singularities in the source term by the use of narrow Gaussian
distributions. We have been able to reproduce earlier results for
equatorial circular orbits that were computed using the frequency
domain formalism. The time domain approach is however geared for a
more general evolution, for instance of nearly geodesic orbits under
the effects of radiation reaction.
\end{abstract}


\maketitle

\section{Introduction}

It will not be long before the current set of interferometric
gravitational wave detectors such as LIGO, VIRGO, GEO and TAMA achieve
sensitivities at which detections are expected to
be possible with a reasonable event rate. The ground-based
interferometric detectors are now starting full-blown ``science runs"
and the plans for more sensitive second and even third generation
detectors are in the drawing boards. Even LISA, the space-based
observatory, seems to be a possibility in a reasonably near future.

This work tries to help in setting the stage for realistic
calculations of what now seems to be a very likely occurrence in
active galactic nuclei: the capture of compact stellar-sized objects
by the supermassive black holes lurking in the center of most
galaxies. This type of events will generate gravitational waves 
within LISA's planned sensitivity band.

Since the captured objects are small compared to the capturing black
hole, they can be treated in a first approximation by 
black hole perturbation theory. But one thorny problem
still remains that has stalled progress in this area. We can model the
waves coming from many kinds of stationary particle orbits around the
black hole, but real orbits will not be stationary. The emission of
gravitational waves will back-react on the object making the orbit
decay in time. One of the challenges in this area is to find workable
prescriptions for these radiation reaction forces that can be used in
numerical simulations that can produce accurate waveforms and energy
fluxes for observers at large distances from the hole to be used as
templates for the interferometric detectors. The prescriptions that
have been worked out so far involve calculating quite complex
integrals along the past light cone of the
particle \cite{minorr,QW1}. Since almost all perturbative treatments of
particles around rotating holes are done in the frequency domain,
applying these radiation reaction schemes has proved
challenging. Frequency domain computations become computationally
expensive when one is dealing with realistic non-circular orbits.

We try here to work out a simple case of equatorial circular particle
orbits around a rapidly rotating black hole to show how to treat this
problem perturbatively in the time domain. We hope this might help to
open the door to the use of the proposed fully relativistic radiation
reaction schemes in the calculation of the more complex but
interesting problem of modeling the orbital decay and gravitational
radiation emission of particles in more general elliptic and/or
inclined orbits.  Up to now, almost all treatments have been based on
the separability of the equation in the frequency domain and calculate
the energy and waveforms for the first few $\ell$ multipoles of the
spheroidal harmonics expansion once the radial part has been dealt
with.

Using the frequency domain approach, quite a few detailed simulations
of the gravitational waves emitted by a particle in a bound orbit
around a black hole have been carried out in recent years. The first
results for radiation emitted by a particle in orbit around a
Schwarzschild black hole were carried out by Detweiler, who pioneered
these techniques \cite{Detw}. In a series of six papers, Poisson and
various collaborators studied in detail the gravitational wave
emission to infinity and into the horizon of a particle in circular
orbit around a non-rotating hole \cite{Pas,Poi1,Finn,Poi2,Poi3,Poi4}.

Other groups have been doing extensive work on Post-Newtonian
expansions for various types of particle orbits around Kerr holes (see
\cite{Mino} and references therein). Work has been done on eccentric
orbits around Schwarszchild \cite{Tan1}, circular orbits slightly
inclined away from the equatorial plane of Kerr holes \cite{Shiba},
and particles with spin orbiting Kerr holes on equatorial circular
orbits \cite{Suki}.

\begin{table}
  \centering
   \begin{tabular}{|c||c|} \hline
   {\bf Authors} & {\bf Type of result}
        \cr \hline \hline
   Detweiler (1978) & equatorial \cr
   Poisson (1993,1995) &  circular orbits around Kerr BH \cr \hline
   & infall trajectories with \cr
   Kojima \& Nakamura (1984) & ang mom $\neq$ 0 \cr \hline
   & infall along symmetry \cr
   Sasaki \& Nakamura (1982) & axis of Kerr BH \cr \hline
   & spinning particles in \cr
   Tanaka {\it et al} (1996) & circ equatorial orbits\cr \hline
   & slightly eccentric and \cr
   Hughes (2000) & non-equatorial orbits\cr \hline
   \end{tabular}
  \caption{A summary of  some of the more relevant frequency domain results
  fora binary black hole system in the particle limit
  that have appeared in the recent literature.}\label{listah}
\end{table}

For single static orbits (ignoring the decay by radiation reaction) to
get good accuracies in the frequency domain approach involves summing
for values of $\ell$ up to 12 for quite a few of the above-harmonic
frequencies, requiring close to 3000 iterations using a numerical
algorithm that involve complex Green's function
calculations \cite{Hug}. Further discussion of the number of $\ell$ 
modes needed, and also a discussion of the self-force in the case
of a scalar field can be seen in the paper by Burko \cite{Burko}.

\section{Time domain evolution of the Teukolsky equation}

We will perform the numerical evolution with the latest version of
a code originally developed by Laguna and
collaborators \cite{Laguna} to evolve the Teukolsky
equation in the time domain from a generic initial
data slice. The method analyzes the radiation at ``infinity'' by
dealing with the $s = -2$ version of the equation. Then the
equation is rewritten as a first order matrix equation and
numerically integrated. The code has been tested
treating scalar fields, scattering of gravitational waves and
analysis of quasi-normal ringing and power law tails of the
outgoing radiation.

We start with the original Teukolsky equation \cite{Teuk} 
before separation is performed,
\begin{eqnarray}
&&{}\left [ \frac{(r^2+a^2)^2}{\Delta} - a^2 \sin^2 \theta \right ]
\frac{\partial^2 \psi}{\partial t^2} + \frac{4 M a r}{\Delta}
\frac{\partial^2 \psi}{\partial t
\partial \phi} + \left [ \frac{a^2}{\Delta} - \frac{1}{\sin^2 \theta} \right
]
\frac{\partial^2 \psi}{\partial \phi^2}
- \Delta^{-s} \frac{\partial}{\partial r} \left ( \Delta^{s+1}\frac{\partial
\psi}{\partial r}\right ) \nonumber \\
&&
- \frac{1}{\sin \theta} \frac{\partial}{\partial \theta} \, \left ( \sin
\theta \frac{\partial \psi}{\partial \theta} \right )
-2 s \left [\frac{a (r-M)}{\Delta} + \frac{i
\, \cos \theta}{\sin^2 \theta} \right ] \frac{\partial
\psi}{\partial \varphi} - 2 s \left [ \frac{M(a^2 - r^2)}{\Delta} -
r - i\, a\, \cos \theta \right ] \frac{\partial
\psi}{\partial t} \nonumber \\
&& + \left [ s^2 \cot^2 \theta - s \right ] \psi = 4 \pi (r^2+a^2\cos^2\theta) T
\end{eqnarray}
where $\Delta=r^2-2M r+a^2$ postulating an ansatz for the $\psi$
function of the form,
\begin{equation}
  \psi = \Psi(r,t,\theta)\:e^{i m \phi}.
\end{equation}
To match this $\phi$ dependence, we write 
the matter source term as a similar Fourier decomposition
\begin{equation}
  T = T_m(r,t,\theta)\:e^{i m \phi}.
\end{equation}

With a modicum of algebra one can rewrite the  Teukolsky
equation as, 
\begin{equation} \label{mteuk}
- \frac{\partial^2 \Phi}{\partial t^2} - A \frac{\partial
\Phi}{\partial t} + b^2 \frac{\partial^2 \Phi}{\partial r^{*2}} +
\frac{c}{\sin \theta} \frac{\partial}{\partial \theta} \left ( \sin
\theta \frac{\partial \Phi}{\partial \theta} \right ) - V \Phi =
T_f.
\end{equation}

This is possible with the following definitions
\begin{equation}
\Phi = \Psi \sqrt{(r^2 + a^2) \Delta^s},
\end {equation}

\begin{equation}
A = \frac{1}{\Sigma^2} \left ( 2 s \left [ r \Delta - M (r^2 +
a^2)\right ] + i \left [ 4 M a r m + 2 s a \Delta \cos \theta
\right ]\right ),
\end{equation}

\begin{equation}
b^2 = \frac{(r^2 + a^2)^2}{\Sigma^2} \hspace{1.2 in} c =
\frac{\Delta}{\Sigma^2},
\end{equation}

\begin{eqnarray}
&&{}V = \frac{1}{\Sigma^2} \left[ \Delta (s^2 \cot^2 \theta - s) + m^2
(\Delta \sin^{-2} \theta - a^2) + 2 s m \Delta \cot \theta \sin^{-1}
\theta \nonumber\right. \\ &&\left. \hspace*{0.6in} - 2 i s m a r (r - M) \right]  +
f \frac{(r^2 + a^2)}{\Sigma^2 \Delta^s} \frac{\partial^2
f}{\partial r^{*2}}
\end{eqnarray}

\begin{equation}
f^2 = (r^2 + a^2) \Delta^s,
\end{equation}

\begin{equation}
T_f = - \frac{4 \pi (r^2+a^2\cos^2\theta) T f \Delta}{(r^2 + a^2)^2 - a^2 \Delta
\sin \theta}
\end{equation}
where $\Sigma^2=(r^2+a^2)^2-\Delta a^2\sin^2\theta$.

One now introduces an auxiliary field $\Pi$ that converts the
Teukolsky equation to a set of 2 coupled PDEs which are first
order in space and time
\begin{equation}
\Pi = \frac{\partial \Phi}{\partial t} + b \frac{\partial
\Phi}{\partial r^*},
\end{equation}
\begin{equation}
\frac{\partial \Pi}{\partial t} + b
\frac{\partial \Pi}{\partial r^*} = b (A -\frac{\partial
b}{\partial r^*}) \frac{\partial \Phi}{\partial r^*}  +
\frac{c}{\sin \theta} \frac{\partial}{\partial \theta} \left ( \sin
\theta \frac{\partial \Phi}{\partial \theta} \right ) - A \Pi - V
\Phi - T_f.
\end{equation}
The code we are using produces stable evolutions of these equations
using a two-step Lax-Wendroff method.

As in other applications of this Teukolsky code, we impose boundary
conditions at the edges of the polar computational domain, i.e. at the
black hole horizon, at the rotation axis and at the far end of the
radial grid. The condition near the horizon is $\Phi = \Pi = 0$, due
to the known asymptotic behavior of the fields \cite{Laguna,Teuk}. At
the outer boundary we impose usual outgoing boundary conditions. Since the
wave equation has a potential, errors due to reflection from this
boundary bounce back into the computational domain. We  deal with
this by making the computational domain in the radial
direction large enough that any numerical reflections will not 
make it back to the
point where we compute the waveforms and energy flux in the time
allotted for the simulation. At the axis of rotation of the black hole
one imposes either the condition $\Phi = 0$ or $\partial_{\theta}
\Phi = 0$ depending on the parity of the field specified by the
azimuthal integer $m$.

The initial data we take for the fields is zero. Since we are
considering the Teukolsky equation with a source we are therefore
violating the constraints of general relativity on the initial
data. This would correspond to the particle ``appearing suddenly'' and
therefore generates an artificial burst of radiation.  To get more
realistic initial data slices, one could try constructing initial data
sets that correctly satisfy the Hamiltonian and momentum constraints
with prescriptions like that of Bowen and York \cite{BY} in the limit
in which one black hole is very small, for instance, or modifications
of particle limit data sets such as those of Lousto and
Price \cite{PrLou}. These prescriptions specify the $( g_{ij},
K_{ij} )$ sets for use in standard ADM evolutions. To convert these
quantities to the set $(\Phi,
\delta_{t}\Phi)$ needed for the Teukolsky evolution, one can use
the formulae developed in reference \cite{Campa}.  We plan to explore
the extent of the benefits that this more accurate intial data
specification may bring to enhance the orbital simulations in the
particle limit case future work.

The next step is to construct the source term for a general geodesic
trajectory of the particle moving in the background Kerr geometry. To
include the effect of radiation reaction one treats it as a slight
deviation of the background geodesic orbital motion.

In the time domain Teukolsky equation the main source term is
given by
\begin{equation}
T = 2 \rho^{-4} T_4,
\end{equation}
where
{
\begin{eqnarray}
&&{}T_4 = (\Delta + 3 \gamma - \gamma^* + 4 \mu + \mu^*)[(\delta^*
- 2 \tau^* + 2 \alpha)T_{nm^*} \nonumber \\ &&{}- (\Delta + 2
\gamma - 2 \gamma^* + \mu^*)T_{m^*m^*}] + (\delta^* - \tau^* +
\beta^* + 3 \alpha + 4 \pi) \nonumber \\ && \times [(\Delta + 2
\gamma + 2 \mu^*)T_{nm^*} - (\delta^* - \tau^* + 2 \beta^* + 2
\alpha) T_{nn}].
\end{eqnarray}
}

All the differential operators and Newman--Penrose coefficients, as
well as the choice of Kinnersley tetrad vectors follow the
notation of \cite{Teuk}. By choosing their values in
Boyer-Lindquist coordinates and expanding everything we can end up
with an explicit but quite complicated form of the source term
that is amenable for numerical calculations\cite{thes}.

We obtain the $r(t)$, $\theta(t)$, and $\phi(t)$ by discretizing
and integrating the geodesic equations in the Kerr background,
\begin{eqnarray}
\Sigma^2\left({dr\over d\tau}\right)^2 && =\left[E(r^2+a^2) - a
L_z\right]^2- \Delta\left[r^2 + (L_z - a
E)^2 + Q\right]\;,\label{eq:rdot} \\
\Sigma^2\left({d\theta\over d\tau}\right)^2 && = Q - \cot^2\theta
L_z^2 -a^2\cos^2\theta(1 -
E^2)\;,\label{eq:thetadot} \\
\Sigma\left({d\phi\over d\tau}\right) && = \csc^2\theta L_z +
aE\left({r^2+a^2\over\Delta} - 1\right) -
{a^2L_z\over\Delta}\;,\label{eq:phidot} \\
\Sigma\left({dt\over d\tau}\right) && =
E\left[{(r^2+a^2)^2\over\Delta} - a^2\sin^2\theta\right] +
aL_z\left(1 - {r^2+a^2\over\Delta}\right)\;.\label{eq:tdot}
\label{eq:geodesiceqns}
\end{eqnarray}

The source term is singular in those points where the particle is
located at each timestep. We treat this \footnote{An alternate way to
treat the polar angle dependence of the source term is to expand it 
using spin-weighted spherical harmonics and keep the as many terms as 
needed. One must note that this approach 
has limited applicability and is hard to extend for more generic orbits 
(inclined orbits, for example). However, it is very useful for testing 
and verificational purposes. } by using the approximation in which we 
substitute the delta function by a very narrow Gaussian function 
(whose width is only of a few gridpoints) :
\begin{equation}\label{gauss}
\delta(x - x(t))\,\approx\,\frac{1}{\sqrt{2
\pi}\,\sigma}\,\exp\left(\frac{-(x-x(t))^2}{2
\sigma^2}\right) \hspace{0.3 in}{\rm for}\;\sigma \;{\rm small}.
\end{equation}

One must show that this substitution is adequate by showing
convergence of the obtained waveform as $\sigma\,\rightarrow\,0$, and
one must take special care in normalizing the mass correctly as the
particle moves in the background Kerr spacetime so that integration
over the whole Gaussian density profile gives a constant mass
throughout the whole trajectory and to satisfy $\nabla_{\mu} T^{\mu
\nu} = 0$. This translates in that the Gaussian must approximately
integrates to one on the spatial manifold. To achieve this with the
correct three-metric of the background black hole spacetime, we
normalize the mass-density of the particle by the factor,
\begin{equation}\label{normlz}
  N = \frac{\sqrt{\gamma^{(3)}}}{\sqrt{g^{(3)}}},
\end{equation}
where $\gamma^{(3)}$ is the 3-metric of flat space and $g^{(3)}$
is the 3-metric of a slice of constant Boyer-Lindquist time of the
Kerr spacetime. We then normalize the particle mass as it moves
thru the space by multiplying the quantity $\mu$ in the source
term expression by this factor $N$. We handle the Dirac deltas in
the radial and angular direction through the Gaussian approximation,
whereas the $\delta(\phi - \phi(t))$
function we handled analytically. The normalizations presented are
common in the SPH literature \cite{lagunasph}.

\section{Numerical Results}

We now want to calculate the energy flux emitted in the form of
gravitational waves from the binary that an orbiting observatory
such as LISA would be able to detect. In this paper we focus on
calculating fluxes from circular equatorial orbits without the
effects of radiation reaction mainly to test the viability of this
approach and to show stability, convergence, and agreement with
previous results of high accuracy frequency domain
calculations\cite{samkip}. In the next section we will explain the
advantages of working in the time domain for future calculations
of eccentric and inclined particle orbits.

The code monitors the value of $\psi$ at a set point in the grid
far from the horizon and computes the energy and angular momentum
fluxes in gravitational radiation according to the formulas \cite{
Gaur, djdt1} :
\begin{eqnarray}
\frac{{\rm d} E}{{\rm d} t}&=&\lim_{r\to\infty}\left\{\frac{1}{4
\pi r^6} \int_{\Omega} {\rm d}\Omega \left| \int_{-\infty}^{t}
\,{\rm
d}\tilde{t}\,\psi(\tilde{t},r,\theta,\varphi)\right|^{2}\right\} \\
\frac{{\rm d} L_z}{{\rm d}t}&=&-\lim_{r\to\infty}\left\{\frac{1}{4
\pi r^6}\mathbf{Re}\left[\int_{\Omega} {\rm d}\Omega
\left(\partial_{\varphi}\,\int_{-\infty}^{t} \,{\rm
d}\tilde{t}\,\psi(\tilde{t},r,\theta,\varphi)\right)\right.\right.\nonumber\\
&&\left.\left.\times\left(\int_{-\infty}^{t} \,{\rm d}t'\,
\int_{-\infty}^{t'} \,{\rm
d}\tilde{t}\,\bar{\psi}(\tilde{t},r,\theta,\varphi)\right)\right]\right\}\;
\;\;\;\;\, {\rm d}\Omega=\sin\theta{\rm d}\theta{\rm d}\varphi.
\label{jfl}
\end{eqnarray}

The code produces output files listing (a) the real and imaginary
parts of the $\psi$ field at representative locations of the 2-D
grid, and (b) the computed energy and angular momentum fluxes for
different values of $m$. And all this is done periodically at
preselected times along the length of the simulation for a
circular orbital geodesic and from here the fluxes are postprocessed.

\begin{figure}\label{cfigure}
\epsfysize=80mm \epsfbox{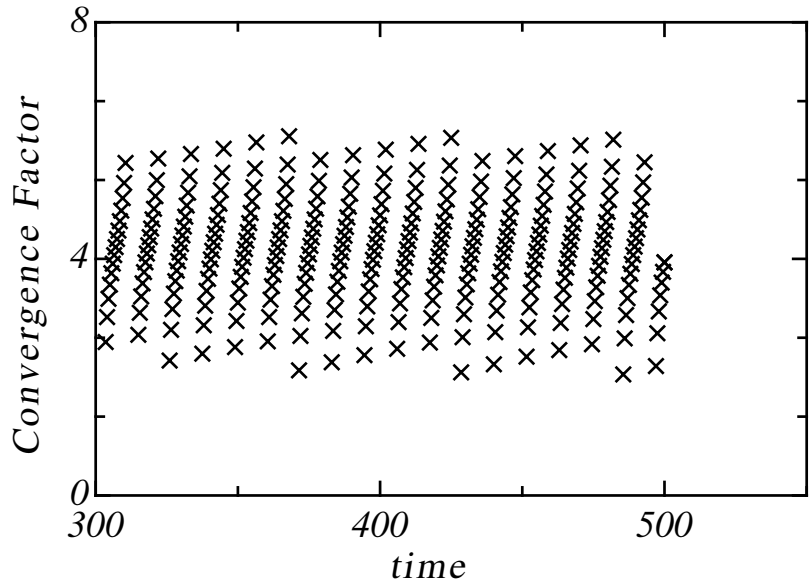} \caption{Analysis of
convergence factor $\mathcal{C}$ calculated from evolved data at a
point at a radius of  $100 M$ in the equatorial plane of the black
hole. Results show that the code has second order convergence (factor=4). 
Oscillations in the convergence factor for the fields measured at
a point is common in systems where the solution itself oscillates. 
One can get a smoother plot by showing convergence of quantities
in L2 norm, for example.}
\end{figure}

We made standard convergence tests by assuming that the numerical
values included errors that decrease as a power of the grid
spacing $h$. Specifically we assume that the value of the
Teukolsky function $\Phi$ at any point in the computational domain
goes like
\begin{equation}\label{errpsi}
  \Phi_{(h)} = \Phi_{exact} + C h^{a} +  \mathbf{O}(h^{a+1}).
\end{equation}
Grids typically used in this work are $-40M \leq r_i^* \leq 460M$ and
$0 \leq \theta_j \leq \pi$ and $i \simeq 6400$ and $j \simeq 60$.
Therefore, if we run at various resolutions (L: 2000X20, M: 4000X40,
H: 8000X80) we can get a quantity $\mathcal{C}$ that measures the
order of convergence {\it a} of the code. This is defined as,
\begin{equation}\label{confac}
  \mathcal{C} = \frac{\Phi_{L} - \Phi_{M}}{\Phi_{M} -
\Phi_{H}},
\end{equation}
and results for it are depicted in figure 1 where we see second order
convergence of the evolution code. We also verified that the results
are independent of the Gaussian width $\sigma$ for both the radial and
angular Gaussian distributions used to approximate the location of the
particle in the discrete spatial grid (see figure 5).

Since equatorial orbits have been treated with great
precision in the frequency domain method, it is important to show
that our method gives comparable results if one looks at
quantities like the average gravitational energy flux at large
distances from the horizon. Finn and Thorne \cite{samkip}
published a comprehensive work that tabulates and includes all
previous results for high-precision gravitational wave flux and
wave amplitudes coming from many kinds of representative compact
objects in equatorial circular orbits around massive central black
holes. In Table
\ref{compft} we present representative comparisons with the results
of formulas (3.8) and (3.10) of that paper using the precise
relativistic corrections included in Tables III - VI that they
also provide. 

\begin{table}
  \centering
   \begin{tabular}{|c||c||c|} \hline
   {\bf m mode} & {\bf Time domain energy flux} & {\bf Finn-Thorne flux }
\cr
    \hline \hline
    && \cr
    $1$ &      $2.5\times10^{-10}$ &  $2.2\times10^{-10}$ \cr \hline
    && \cr
    $2$ &      $1.0\times10^{-07}$ &  $1.4\times10^{-07}$ \cr \hline
    && \cr
    $3$ &      $3.9\times10^{-08}$ &  $3.1\times10^{-08}$ \cr \hline
    && \cr
    $4$ &      $1.0\times10^{-08}$ &  $8.0\times10^{-09}$ \cr \hline
   \end{tabular}
  \caption[Comparisons of gravitational wave energy fluxes for frequency
domain methods and our method.]{Comparisons
  of gravitational wave energy fluxes detected at infinity
  using the frequency domain solution of the Teukolsky-Sasaki-Nakamura
equation
  $\dot{E}_{\infty m}$ as calculated in Ref.\cite{samkip} with the results
measured
  numerically  at $r = 100 M$
while
  evolving the Teukolsky equation in the time domain under the "particle-as
  -a-gaussian-distribution" approximation discussed here. The particle here is in a
circular, equatorial orbit of radius $2.0\, r_{isco}$ about a
black hole with Kerr parameter $a/M = 0.9$ and $\mu/M=0.01$.}\label{compft}
\end{table}

\section{Discussion}

We have shown that it is possible to integrate the Teukolsky equation
in the time domain for the problem of a particle orbiting around a
black hole with the particle was represented as a Gaussian and obtain
results reasonably close to those of the frequency-domain
calculations. The time domain approach is expected to become more
useful for the case of more complicated orbits. We plan in future
papers to discuss how to how to use better approximations to the
Gaussian like those of reference \cite{PrLo2}, and how to incorporate
radiation-reaction effects. Also, the small discrepancies with the
Thorne--Finn results need to be further investigated to define well a
``window of reliability'' for the results of the code. Another element
to be incorporated is to take into account the radiation flux falling
into the black hole, since this will also influence trajectories
corrected by radiation-reaction. The first approach will be to correct
the orbit trajectory using energy-balance arguments, as for instance
in
\cite{Hug,Bur2}. Further possibilities include incorporating proposed 
radiation reaction forces directly into the integration of the orbit.

\begin{figure}
\epsfysize=80mm \epsfbox{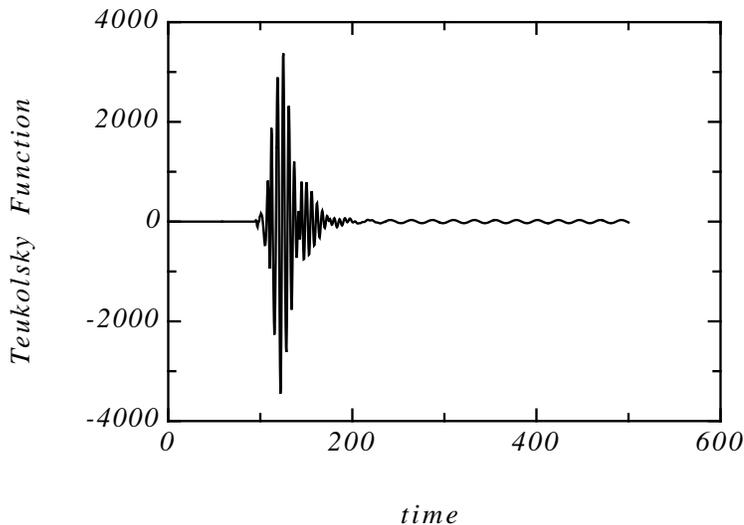} \caption{Representative
data showing the evolution of the real part of the Teukolsky function
$\Phi$ for the $m=3$ mode of a particle orbiting a rapidly
rotating Kerr hole of $a = 0.9 M$ at a radius of $2\, r_{isco}$ 
($r_{isco}=2.32 M$) at an observation point $r/M=100$, $\theta=\pi/2$.}
\end{figure}

\begin{figure}
\epsfysize=80mm \epsfbox{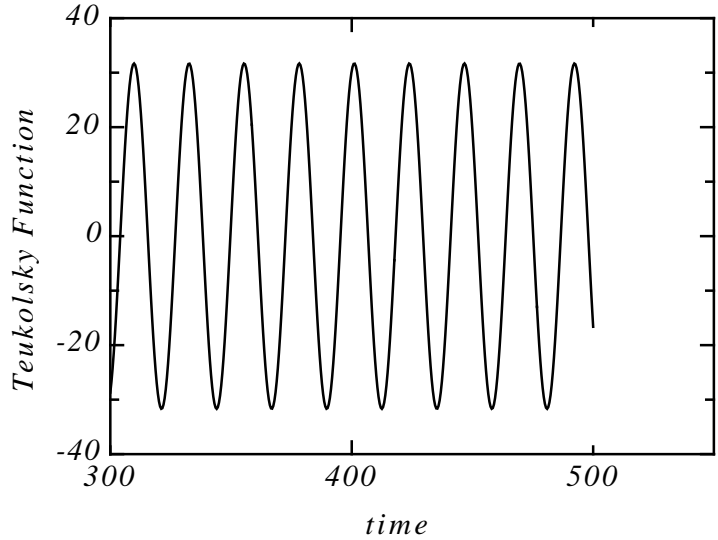} \caption{More detail of
the evolution of the Teukolsky function for the same simulation
showing the stability of the evolution after the system has
settled during a few static cirbular orbits.}
\end{figure}

\begin{figure}
\epsfysize=80mm \epsfbox{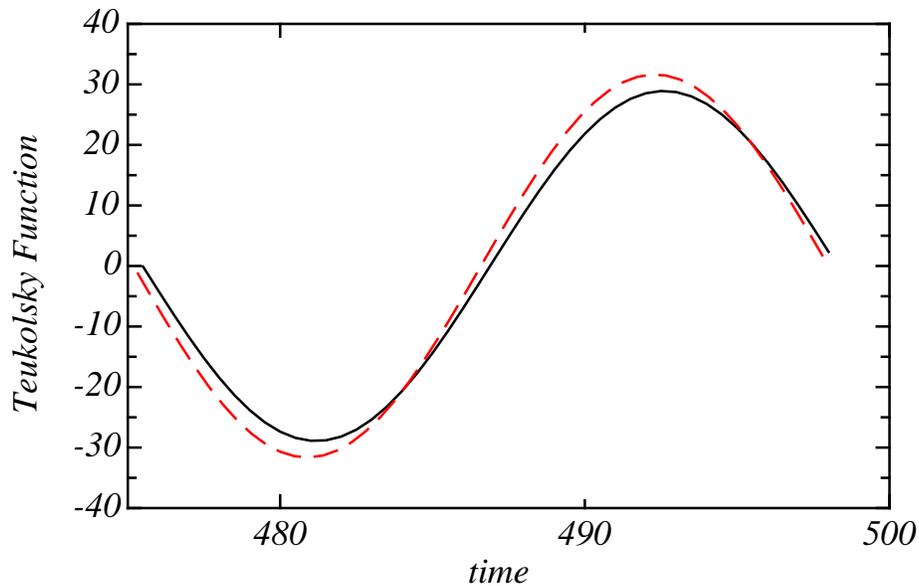} \caption{Teukolsky function 
insensitivity to the radial and angular width of the gaussian distribution 
that models the particle in orbit. The parameters for this 
evolution were the same as used for the other figures. The solid curve
corresponds to radial and angular widths of $0.75 M$ and $0.47$ rad 
respectively, while dotted curve corresponds to widths of $0.25 M$ and 
$0.16$ rad. }
\end{figure}

\begin{figure}
\epsfysize=80mm \epsfbox{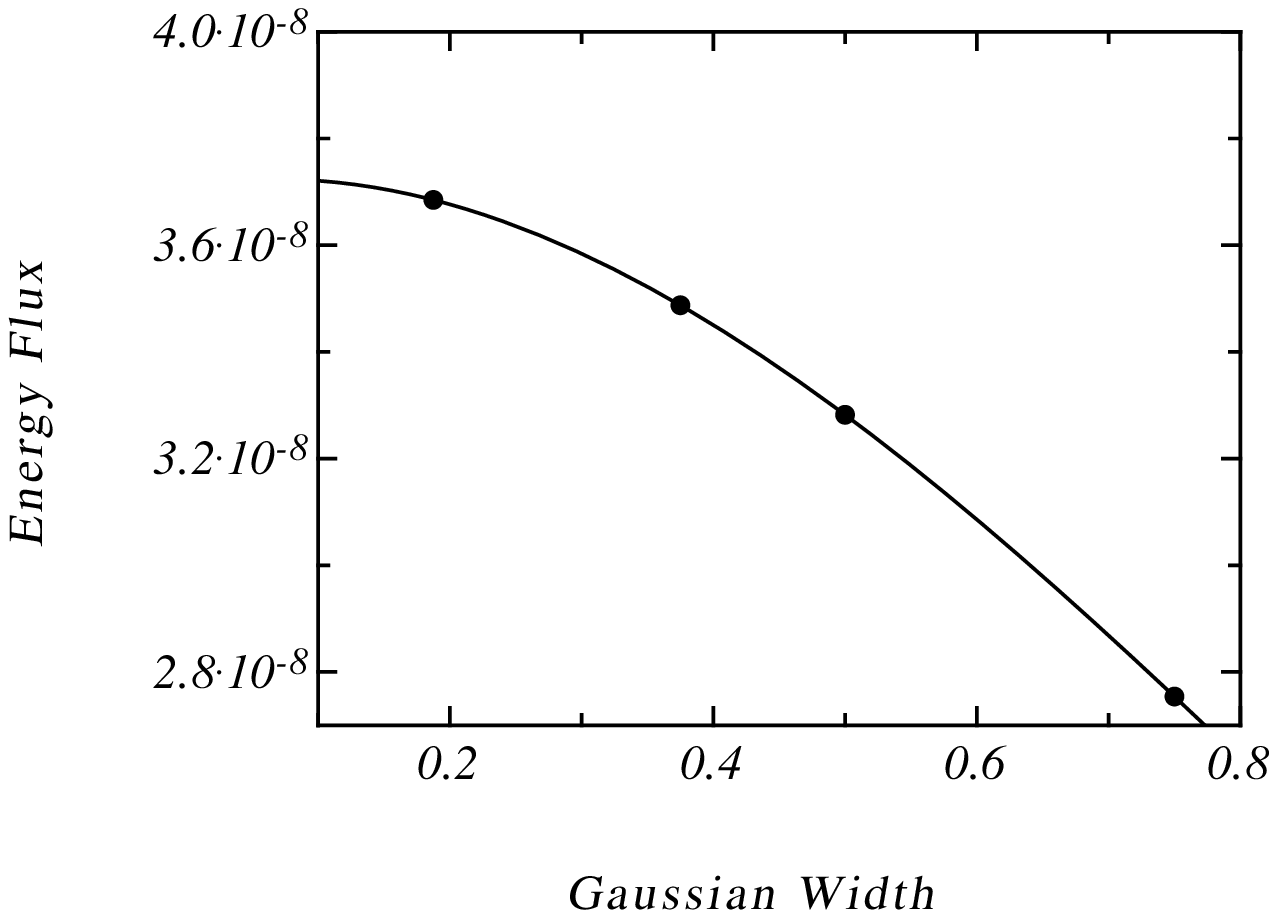} \caption{Convergence of the
code when one makes the Gaussian narrower. Of course one cannot expect
convergence in the traditional sense when one makes the Gaussian 
narrower, since at some point it will become too narrow for the finite
resolution of the code to model it. What can be expected, and is shown
in the figure is that the results tend to a constant value for a while
when the Gaussian is made narrower, and shortly before it becomes too
narrow for the code to resolve. }
\end{figure}

\acknowledgments

We are grateful to Pablo Laguna who continually provided help with the
use of his Teukolsky code during this work and for comments on the
manuscript.  We also thank Eric Poisson for providing results for
comparison and comments. We are indebted to Lior Burko, Scott Hughes,
Carlos Lousto and Karl Martel for comments on the manuscript. RLA is
grateful for support in form of a FIPI ( Institutional Fund for
Starting Research) grant of the Office of the Dean of Research and
Graduate Studies of the Rio Piedras campus at the University of Puerto
Rico. GK acknowledges research support from Long Island
University. This work was supported by NSF grants PHY-0140236,
PHY-9800973 and by funds of the Horace Hearne Institute for
Theoretical Physics at LSU.

\end{document}